\documentclass[a4paper,12pt]{elsarticle}

\usepackage[utf8]{inputenc}
\usepackage[T1]{fontenc}
\usepackage{microtype} 
\usepackage[english]{babel}
\usepackage[utf8]{inputenx} 
\usepackage{hyperref}
\usepackage{appendix}
\usepackage{color}
\usepackage{xcolor}
\usepackage{amsmath}
\usepackage{amssymb}
\usepackage{varioref}
\usepackage{wasysym}
\usepackage{verbatim}
\usepackage{lipsum}
\usepackage{xcolor, tikz, pgfplots}
\usepackage{geometry}
\usetikzlibrary{matrix,calc,positioning,decorations.markings,decorations.pathmorphing,decorations.pathreplacing
}
\usetikzlibrary{arrows,cd,shapes}

\journal{Communications in Theoretical Physics} 

\begin{document}

\begin{frontmatter}


\title{Algebro-geometrical orientifold and IR dualities}

\author[label1,label2]{Federico Manzoni\fnref{1}}

\address[label1]{Mathematics and Physics department, Roma Tre, Via della Vasca Navale 84, Rome, Italy}
\address[label2]{INFN Roma Tre Section, Physics department, Via della Vasca Navale 84, Rome, Italy}

\fntext[1]{federico.manzoni@uniroma3.it}

\begin{abstract}
Orientifold projections are an important ingredient in geometrical engineering of Quantum Field Theory. However, an orientifold can break down the superconformal symmetry and no new superconformal fixed points are admitted (scenario II); nevertheless, in some cases, dubbed scenarios I and III orientifold, a new IR fixed point is achieved and, for scenario III examples, some still not fully understood IR duality seems to emerge. Here we give an algebro-geometrical point of view of orientifold for toric varieties and we propose the existence of relevant operators that deform the starting oriented CFT triggering a flow. We briefly discuss a possible holographic description of this flow.
\end{abstract}

\begin{keyword}
Non-perturbative orientifold \sep RG flow \sep Toric geometry \sep IR duality.
\end{keyword}

\end{frontmatter}


\tableofcontents

\section{Introduction}
AdS/CFT correspondence is a masterpiece of modern physics: its impact in linking physics and geometry has opened new doors and laid the foundations for new research fields. The original Maldacena AdS/CFT correspondence \cite{Maldacena:1997re} can be extended to phenomenologically interesting constructions that require $4D$ SUSY gauge theories with $\mathcal{N}=1$ or $\mathcal{N}=2$. The starting point is to embed our stack of $N$ D$3$-branes into a background space-time of the form $\mathbb{M}^{3,1} \times C(X_5)$ where $C(X_5)$ is a Calabi-Yau (CY) cone over a Sasaki-Einstein (SE) manifold $X_5$. The geometry properties of the CY define the field theory on the branes but for general Calabi-Yau cone it is hard to build up the worldvolume gauge theory. We restrict ourselves to the class of toric CY \cite{Closset:2009sv,1999sd,fulton} since this type of varieties allows an algorithmic construction of the field theories. The first example of this kind of AdS/CFT extensions is due to Klebanov and Witten in their 1998 pioneering work \cite{1998ks}. The toric condition simplifies a lot the geometric description of the Calabi-Yau cone, indeed all the information is stored into the toric diagram, as we will review in the Paragraph \ref{2}, and the emerging field theories can be studied using well established brane tiling techniques \cite{Franco_2006,Franco:2005sm,Franco:2013ana,Yamazaki_2008,hanany2005dimer,Hanany_2012,Hanany:2012hi}.

In Superconformal Quantum Field Theories (SQFTs), that emerge from the geometrical engineering with toric CY cone, we can consider the so-called orientifold projections \cite{Dudas:2006bj,Bianchi:1990yu,Sagnotti:1987tw} that are reviewed in Paragraph \ref{2a}. The action in the string theory side is to make open oriented string unoriented and, according to a Chan-Paton-like analysis, real group such as $SO(N)$ and $Sp(2N)$ are allowed in the emerging SQFTs. In 1995 Polchinski realized \cite{1995pol} that orientifold projections have a simple and elegant interpretation from the point of view of the background space-time: they correspond to not dynamical, mirror-like objects called orientifold planes, defined by $T$-duality as fixed points of the orientifold projections. On the field theory side, orientifold projection affects fields and superpotential, and its effect can be understood using the brane tiling picture \cite{2007do}. 

Let us indicate orientifold with $\Omega$, therefore quantities after orientifold projection will be labelled with the apex $\Omega$. The theory before orientifold is often referred to as the "parent", "oriented", "pre-orientifold" or "unorientifolded" theory while the one after orientifold as "daughter", "unoriented", "post-orientifold" or "orientifolded" theory. Similar names hold for quantities in the theories.\\
The presence of orientifold planes modifies the Renormalization Group (RG) flow, and two different scenarios are mostly investigated in literature. On the one hand, in the scenario I there is a new fixed point and the $R$-charges of operators that are not projected out are the same as the charges of the corresponding parent theory in the large $N$ limit; this gives us a central charge $a^{\Omega}$ that is half the central charge of the parent theory. On the other hand, in the scenario II the daughter theory does not have a fixed point.

Despite this, recently \cite{2020in,2022su}, a new possibility seems to enter in the game: the so-called scenario III. 
In this case it is possible \cite{2020in,2022su} to construct an orientifold projection such that the central charge $a^{\Omega}$ is less than half of the central charge $a$ of the parent theory. Moreover, the $R$-charges before and after the orientifold are not the same. The interesting behavior is that the values of the $R$-charges and central charge coincide, beyond the large $N$ limit, with those of another theory that has the scenario I orientifold. This seems to suggest an IR duality which is in the IR regime because the central charge $a$ gives us, due to $c$-theorem in $4D$, an RG flow ordering of theories. Therefore we know that the scenario I and III orientifolds are lowering the energy moving the theories into their IR regimes.\\

Here we present an algebraic-geometrical interpretation of orientifold, motivated by Greene's flop transition and the search for a quantum geometry \cite{Greene:1996cy,Aspinwall:1993nu,Greene:1995hu,Witten:1993yc}, as morphisms between algebraic varieties. Orientifolds are thought as maps that act both on the states of the theory and on the Calabi-Yau cones geometry. The net effect is an RG flow towards the IR regime, from the original oriented model to an unoriented one (see Figure \ref{fig7} to have a schematic picture in mind).\\
In Paragraph \ref{2} we review toric varieties and toric CY discussing the construction of manifolds as algebraic subvariety of the complex space while in Paragraph \ref{2a} we review orientifold projection and its possible scenarios. In Sections \ref{3} and \ref{cf} we discuss the algebro-geometrical interpretation of orientifold and the possible way RG flows triggered by the orientifold itself can be holographically interpreted.
\section{Background material: toric geometry and orientifold}

\subsection{Toric varieties and toric Calabi-Yau}\label{2}
Toric geometry is a branch of algebraic geometry and a toric variety is, by definition, an algebraic variety containing an algebraic torus as an open dense subset, such that the action of the torus on itself extends to the whole variety. To be more specific, an $n$-dimensional toric variety $\mathcal{M}$ has an algebraic torus action in the sense that the algebraic torus $\mathbb{T}^n=(\mathbb{C}^*)^n$ is a dense open subset and there is an action $\mathbb{T}^n \times \mathcal{M} \rightarrow \mathcal{M}$. The greatest point in favor of toric geometry is that geometry of a toric variety is fully determined by combinatorics.\\
In the case of a toric Calabi-Yau variety the information about the geometry is summarized in the so-called toric diagram which is a polytope embedded in a $\mathbb{Z}^{n-1}$-lattice. Moreover, when we consider a toric Calabi-Yau threefold cone the action of the algebraic torus enlarges the isometry group of the variety from $U(1)$, the action induced by the Reeb vector field, to $U(1)^3$. Let us briefly review some possible constructions of toric varieties \cite{Closset:2009sv,1999sd,fulton}.
\subsubsection{Homogeneous coordinates approach to toric varieties}\label{hca}
The simplest way we can image toric variety is as generalization of weighed projective space \cite{cox}. Let us recall first the definition of the $(m-1)$-dimensional weighed projective space
\begin{equation}
   \mathbb{CP}^{m-1}= \frac{\mathbb{C}^m \setminus \{0\}}{\mathbb{C}^*},
\label{wps}
\end{equation}where the quotient by $\mathbb{C}^*$ is taken into account by the identification $(z_1,...,z_{m}) \sim (\lambda^{i_1}z_1,...,\lambda^{i_{m}}z_{m})$ where $\lambda \in \mathbb{C}^*$ and $({i_{1}},...,{i_{m}})$ are the so-called coordinates weights. An $n$-dimensional toric variety $\mathcal{M}$ is the generalization where we quotient by more than one $\mathbb{C}^*$ action and the set that we subtract is a subset $U_{\Sigma}$ which contains not only the origin
\begin{equation}
    \mathcal{M}=\frac{\mathbb{C}^m \setminus \{U_{\Sigma}\}}{(\mathbb{C}^*)^{m-n} \times \Gamma},
\label{toric1}
\end{equation}
where $\Gamma$ is an abelian group. This variety has an algebraic torus action given by $(\mathbb{C}^*)^{m-m+n}=(\mathbb{C}^*)^{n}$. Definition (\ref{toric1}) emerges in the context of cones and fans that we are going to discuss.\\
Let $M$ and $N$ be dual $n$-dimensional lattices isomorphic to $\mathbb{Z}^n$ and consider the vector spaces $ M_{\mathbb{R}}$ and $N_{\mathbb{R}}$ to be the subspace of $\mathbb{R}^n$ spanned, respectively, by vectors in $M$ and $N$. We define the Strongly Convex Rational Polyhedral Cone (SCRPC) $\sigma \in N_{\mathbb{R}} \subset \mathbb{R}^n$ as the set
\begin{equation}
    \sigma:=\bigg\{\sum_{i=1}^m a_i\Vec{v}_i \bigg| a_i \in \mathbb{R}, a_i\geq 0, \Vec{v}_i \in \mathbb{Z}^n \  \forall i\bigg\}
\end{equation}
for a finite number of vectors $v_i$ and satisfying the condition $(\sigma) \cap (-\sigma)=\{0\}$. Let us analyze this definition: we consider an $n$-dimensional lattice $N \simeq \mathbb{Z}^n$, a SCRPC is an $n$ or lower dimensional cone in $N_{\mathbb{R}} $ with the origin of the lattice as its apex, bounded by hyperplanes (polyhedral) with its edges spanned by lattice vectors (rational) and such that it does not contain complete lines (strongly convex). The dimension of a SCRPC $\sigma$ is the dimension of the smallest subspace of $\mathbb{R}^n$ containing $\sigma$. Two important concepts are useful in SCRPC theory:
\begin{itemize}
    \item edges: these are the one dimensional faces of $\sigma$, the vectors $\Vec{g}$ associated to the edges are the generators of $\sigma$;
    \item facets: these are the codimension one faces.
\end{itemize}
\begin{figure}[h]
    \centering
    {\includegraphics[width=.22\textwidth]{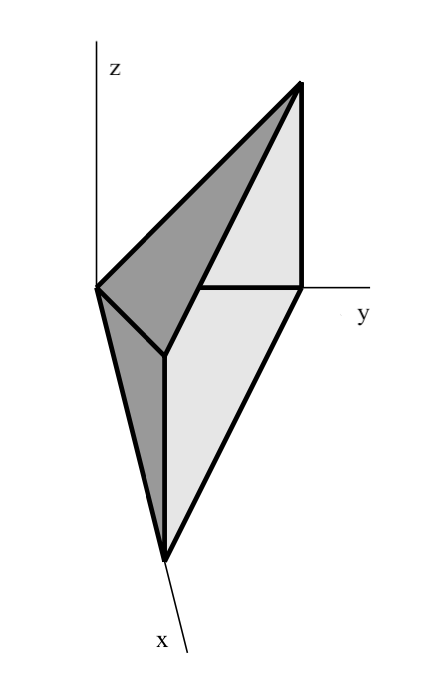}}
    \label{fig18a}\quad
    {\includegraphics[width=.35\textwidth]{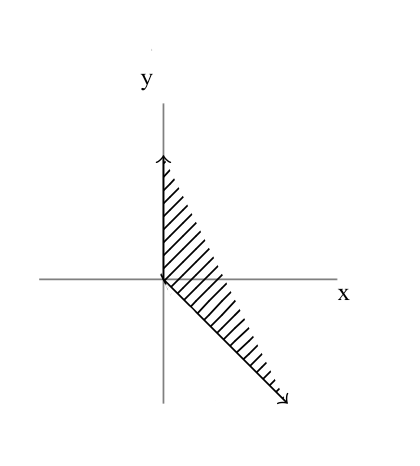}
    \label{fig18b}}
\caption{\textit{Examples of SCRPCs. Left: SCRPC in $\mathbb{R}^3$, its one dimensional faces are identified by the vectors $\Vec{g}_1=(1,0,0), \Vec{g}_2=(0,1,0),  \Vec{g}_3=(0,1,1),  \Vec{g}_4=(1,0,1)$ in $\mathbb{Z}^3$. Right: SCRPC in $\mathbb{R}^2$, its one dimensional faces are identified by the vectors $\Vec{g}_1=(0,1), \Vec{g}_2=(1,-1)$ in $\mathbb{Z}^2$.}}
\label{conespoli}
\end{figure}
A collection $\Sigma$  of SCRPCs in $N_{\mathbb{R}}$ is called fan if each face of a SCRPC in $\Sigma$ is also a SCRPC in $\Sigma$ and the intersection of two SCRPCs in $\Sigma$ is a face of each. Examples of SCRPCs and fan are reported in Figure \ref{conespoli} and Figure \ref{fig19}.
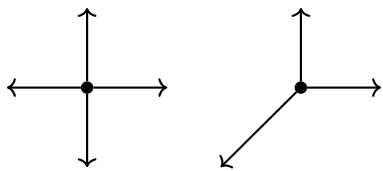
\begin{figure}[ht]
\centering{\begin{tikzpicture}[ auto, scale=0.6] 
    \node [] (0) at (0,2) {};
    \node [circle, draw=black , fill=black, inner sep=0pt, minimum size=1.5mm] (1) at (0,0) {};
    \node [] (2) at (2,0) {};
    \node [] (3) at (-2,0) {};
    \node [] (4) at (0,-2) {};
    \draw (1) to  [] (0) [->, thick]; 
    \draw (1) to  [] (3) [->, thick];
    \draw (1) to  [] (2) [->, thick];
    \draw (1) to  [] (4) [->, thick];
 
\end{tikzpicture}} 
\centering{\begin{tikzpicture}[ auto, scale=0.6] 
    \node [] (0) at (0,2) {};
    \node [circle, draw=black , fill=black, inner sep=0pt, minimum size=1.5mm] (1) at (0,0) {};
    \node [] (2) at (2,0) {};
    \node [] (3) at (-2,-2) {};
    \draw (1) to  [] (0) [->, thick]; 
    \draw (1) to  [] (3) [->, thick];
    \draw (1) to  [] (2) [->, thick];
\end{tikzpicture}} 
\caption{\textit{Examples of fan. Left: fan of $\mathbb{CP}^1 \times \mathbb{CP}^1$, we have four one dimensional SCRPCs (the vectors) and four two dimensional SCRPCs (the quadrants). Right: fan of $\mathbb{CP}^2$, we have three one dimensional SCRPCs (the vectors) and three two dimensional ones (the trians).}}
\label{fig19}
\end{figure}\\
\\
\\
\\

Let us consider a fan $\Sigma$ and we call $\Sigma_{1d}$ the set of one dimensional SCRPCs; let $\Vec{v}_i$ with $i=1,...,m$ be the whole set of vectors generating the one dimensional SCRPCs in $\Sigma_{1d}$\footnote{Note that obviously $m$ is equal to the number of one dimensional cone and so to the number of elements in $\Sigma_{1d}$}. To each vector $\Vec{v}_i$ we associate a homogeneous coordinate $z_i \in \mathbb{C}$ and we define the set 
\begin{equation}
    U_{\Sigma}:= \bigcup_I\{(z_1,...,z_m)|z_i=0 \ \forall i \in I\},
\end{equation}
where the union is taken over all the sets having $I \subseteq \{1,...,m\}$ for which $z_i$ with $i \in I$ does not belong to a SCRPC in $\Sigma$. 

At this point we need to discuss how the $(\mathbb{C}^*)^{m-n} \times \Gamma$ acts on $\mathbb{C}^m$. First of all, let us clarify the nature of the abelian group $\Gamma$: this is given by 
\begin{equation}
    \Gamma:=\frac{N}{\tilde{N}},
\end{equation}
where $\Tilde{N} \subseteq N$ is the sublattice generated over $\mathbb{Z}^n$ by the vectors $\Vec{v}_i$. In other words, vectors $\Vec{v}_i$ not necessarily generate all $N$, in general they generate only $\tilde{N}$; on the one hand, if $\tilde{N}=N$ then $\Gamma$ is trivial and it does not play any role. On the other hand, if $\Gamma$ is no trivial our variety develops orbifold singularity. \\
Let us now discuss the algebraic torus $(\mathbb{C}^*)^{m-n}$ action. Consider the $n \times m$ matrix build up considering the $m$ vectors $\Vec{v}_i$ with $n$ components
\begin{equation}
V_i^k=\begin{pmatrix}
v_1^1 & v_2^1 & \dots & v_m^1 \\
v_1^2 & \ddots & \dots & v_m^2\\
\vdots & \vdots & \ddots & \vdots\\
v_1^n & v_2^n &\dots & v_m^n
\end{pmatrix},
\end{equation}
this induces a map $\phi: \mathbb{C}^m \rightarrow \mathbb{C}^n$ defined by
\begin{equation}
    (z_1,...,z_m) \mapsto \bigg(\prod_{i=1}^m z_i^{v_i^1},...,\prod_{i=1}^m z_i^{v_i^n}\bigg).
    \label{map32}
\end{equation}
Thanks to the rank-nullity theorem\footnote{Given $T$ a linear map between two finite dimensional vector spaces $A$ and $B$ we have that $dim(Im(T))+dim(Ker(T))=Rank(V)+Null(V)=dim(A)$. In our case we know that $\Vec{v}_i$ with $i=1,...,m$ are $m$ linearly independent vectors so $Rank(V)=dim(Im(T))=m$.} the dimension of the Kernel of map (\ref{map32}) must be $m-n$; we can identify it with $(\mathbb{C}^*)^{m-n}$. It is now simple to see how $(\mathbb{C}^*)^{m-n}$ acts on $\mathbb{C}^m$: each $\mathbb{C}^*$ action  is taken into account by
\begin{equation}
    (z_1,...,z_m) \mapsto (\lambda^{Q_1^a}z_1,...,\lambda^{Q_m^a}z_m),
    \label{map33}
\end{equation}
with $\lambda \in \mathbb{C}^*$. We have $m-n$ actions like (\ref{map33}), where for each $a=1,...,m-n$ the charge vectors $Q^a=(Q_1^a,...,Q_m^a)$ belong to the Kernel of the map $\phi$ and therefore must satisfy $m-n$ relations
\begin{equation}
    \sum_{i=1}^m V^k_iQ_i^a=\Vec{0}.
    \label{4.8}
\end{equation}
Quotient $(\mathbb{C}^*)^{m-n}$ out means taking into account the equivalence relations
\begin{equation}
    (z_1,...,z_m) \sim (\lambda^{Q_1^a}z_1,...,\lambda^{Q_m^a}z_m)
    \label{4.9}
\end{equation}
for $a=1,...,m-n$. \\
To summarize, we can define the toric variety as 
\begin{equation}
     \mathcal{M}=\frac{\mathbb{C}^m \setminus \{U_{\Sigma}\}}{(\mathbb{C}^*)^{m-n} \times \Gamma};
\end{equation}
this is an $n$-dimensional variety, with a residual $(\mathbb{C}^*)^n \simeq U(1)^n$ action and the $(\mathbb{C}^*)^{m-n}$ action is quotient out by $m-n$ relations (\ref{4.9}) with weights that satisfy relations (\ref{4.8}). Let us give an example.
\subsubsection*{Example: $\mathbb{CP}^2$}
Let us consider the right fan in Figure \ref{fig19}: this is a $\mathbb{Z}^2$ lattice; we have the three vectors $\Vec{v}_1=(0,1), \Vec{v}_2=(1,0), \Vec{v}_3=(-1,-1)$ that generate one dimensional cones. We have 
three homogeneous coordinates $(z_1,z_2,z_3) \in \mathbb{C}^3$ and the set $U_{\Sigma}$ is given by the origin. Since $m=3$ and $n=2$ we must have one $\mathbb{C}^*$ action and we can find the weights using (\ref{4.8}):
\begin{equation}
\begin{aligned}
  &\Vec{v}_1Q_1+\Vec{v}_2Q_2+\Vec{v}_3Q_3= (0,1)Q_1+(1,0)Q_2+(-1,-1)Q_3=\\
  &=(Q_2-Q_3,Q_1-Q_3)=\Vec{0} \Rightarrow Q_1=Q_2=Q_3=1;
\end{aligned}
\end{equation}
$\mathbb{C}^*$ action is taken into account by the equivalence relation $(z_1,z_2,z_3) \sim \lambda (z_1,z_2,z_3)$. Finally we note that the vectors $\Vec{v}_1$ and $\Vec{v}_2$ generate the whole lattice $\mathbb{Z}^2$ and so $\Gamma$ is trivial. The toric variety corresponds to 
\begin{equation}
    \mathcal{M}=\frac{\mathbb{C}^3 \setminus \{0\}}{\mathbb{C}^*}\equiv \mathbb{CP}^2,
\end{equation}
as we expected.\\

We now give some interesting properties about toric varieties and their fans:
\begin{itemize}
    \item a fan $\Sigma$ is smooth if every SCRPC in $\Sigma$ is smooth, a SCRPC is smooth if is generated by a subset of a basis of $N \simeq \mathbb{Z}^n$;
    \item a fan $\Sigma$ is simplicial if every SCRPC in $\Sigma$ is simplicial, a SCRPC is simplicial if it is generated by a subset of a basis of $\mathbb{R}^n$;
\end{itemize}
These conditions are important since if a fan is smooth the corresponding toric variety also is smooth and if a fan is simplicial then the corresponding toric variety can have at most orbifold singularities. We see immediately that the two spaces described by the fans in Figure \ref{fig19} are smooth since every SCRPC is generated by a subset of a $\mathbb{Z}^2$ basis. An example of toric variety with orbifold singularities is the weighted projective space $\mathbb{CP}_{2,3,1}$; its fan is given by the Figure \ref{fig20} below. Orbifold singularities can be removed by the so-called blow up procedure: roughly speaking, for a $n$-dimensional toric variety we replace the singular locus by $\mathbb{CP}^{n-1}$.
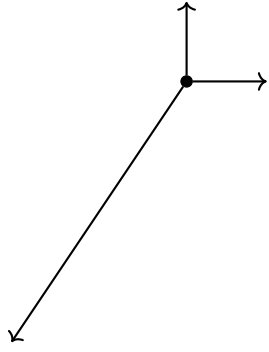
\begin{figure}[ht]
\centering{\begin{tikzpicture}[ auto, scale=0.6] 
    \node [] (0) at (0,2) {};
    \node [circle, draw=black , fill=black, inner sep=0pt, minimum size=1.5mm] (1) at (0,0) {};
    \node [] (2) at (2,0) {};
    \node [] (3) at (-4,-6) {};
    \draw (1) to  [] (0) [->, thick]; 
    \draw (1) to  [] (3) [->, thick];
    \draw (1) to  [] (2) [->, thick];
\end{tikzpicture}} 
\caption{\textit{Fan of $\mathbb{CP}_{2,3,1}$, we have $\Vec{v}_1=(1,0), \Vec{v}_2=(0,1), \Vec{v}_3=(-2,-3)$. It is not smooth but it is simplicial: $\mathbb{CP}_{2,3,1}$ has orbifold singularities.}}
\label{fig20}
\end{figure}

Let us now specialize to the case of our interest: CY threefolds. First of all, we have an $U(1)^3 \simeq (\mathbb{C}^*)^3$ action but there is more: the condition of trivial canonical bundle implies that all the vectors of the fan belong to the same hyperplane, so we can project on this hyperplane obtaining a two dimensional object whose convex hull takes the name of toric diagram. The CY condition implies
\begin{equation}
    \mathrm{CY} \ condition \Rightarrow \sum_{i=1}^m Q_i^a=0 \ \forall a.
\label{CYc}    
\end{equation}
\subsubsection*{Example: the conifold $\mathcal{C}$}\label{excon}
Consider the three dimensional fan given by the four vectors $\Vec{v}_1=(1,0,1), \Vec{v}_2=(0,0,1),\Vec{v}_3=(0,1,1),\Vec{v}_4=(1,1,1)$. We note that these vectors belong to the same hyperplane, so we can project out the third component: we obtain $\Vec{w}_1=(1,0), \Vec{w}_2=(0,0),\Vec{w}_3=(0,1),\Vec{w}_4=(1,1)$. Hence the conifold is a CY variety and its toric diagram is given below.
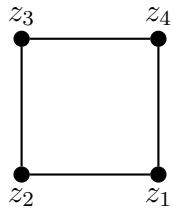
\begin{figure}[h]
\centering{\begin{tikzpicture}[ auto, scale=0.6] 
    \node [circle, draw=black , fill=black, inner sep=0pt, minimum size=2mm] (0) at (0,0) {};
    \node [circle, draw=black , fill=black, inner sep=0pt, minimum size=2mm] (1) at (0,3) {};
    \node [circle, draw=black , fill=black, inner sep=0pt, minimum size=2mm] (2) at (3,0) {};
    \node [circle, draw=black , fill=black, inner sep=0pt, minimum size=2mm] (3) at (3,3) {};
  
    \node  (4) [] at (0,-0.5) {$z_2$};
    \node  (5) [] at (0,3.5) {$z_3$};
    \node  (6) [] at (3,-0.5) {$z_1$};
    \node  (7) [] at (3,3.5) {$z_4$};
  
    \draw (0) to  [] (1) [-, thick]; 
    \draw (1) to  [] (3) [-, thick];
    \draw (3) to  [] (2) [-, thick]; 
    \draw (2) to  [] (0) [-, thick];
\end{tikzpicture}} 
\caption{\textit{Toric diagram of the conifold.}}
\label{fig21}
\end{figure}\\
We have $m-n=4-3=1$ charge vector given by relation (\ref{4.8}):
\begin{equation}
    \begin{pmatrix}
1 & 0 & 0 & 1\\
0 & 0 & 1 & 1\\
1 & 1 & 1 & 1 \\
\end{pmatrix}\begin{pmatrix}
Q_1 \\
Q_2 \\
Q_3 \\
Q_4 \\
\end{pmatrix}=\Vec{0} \Rightarrow \begin{cases}
  Q_1+Q_4&=0 \\
  Q_3+Q_4&=0 \\
   \sum_{i=1}^4 Q_i&=0 
\end{cases}
\end{equation}
and a possible solution is $Q=(1,-1,1,-1)$. As one can note the CY condition is automatically implemented by the fact that the vectors are coplanar. The action of the algebraic torus $\mathbb{C}^*$ is quotient out by the equivalence relation
\begin{equation}
    (z_1, z_2, z_3, z_4) \sim (\lambda z_1, \lambda^{-1} z_2, \lambda z_3, \lambda^{-1} z_4).
    \label{eqrc}
\end{equation}
Let us consider the $\mathbb{C}^*$-invariant polynomials: by (\ref{eqrc}) we note that
\begin{equation}
    Z_1=z_1z_2, \ Z_2=z_1z_4, \ Z_3=z_2z_3, \ Z_4=z_3z_4
\end{equation}
are invariant and this is the minimal basis with which to write all the $\mathbb{C}^*$-invariant polynomials. However, note that these polynomials are not independent but they must satisfy the relation
\begin{equation}
    Z_1Z_4=Z_2Z_3
\end{equation}
which is the defining polynomial of the conifold \cite{1998ks}. 
\subsubsection*{Example: $\frac{\mathbb{C}^2}{\mathbb{Z}_2} \times \mathbb{C}$}\label{exc3}
Consider the fan generated by the vectors $\Vec{v}_1=(0,0,1), \Vec{v}_2=(0,1,1), \Vec{v}_3=(1,0,1), \Vec{v}_4=(-1,0,1) $; projecting out the third component we get $\Vec{w}_1=(0,0), \Vec{w}_2=(0,1), \Vec{w}_3=(1,0), \Vec{w}_4=(-1,0)$ and the toric diagram is 
\begin{figure}[ht]
\centering{\begin{tikzpicture}[ auto, scale=0.6] 
    \node [circle, draw=black , fill=black, inner sep=0pt, minimum size=2mm] (0) at (0,0) {};
    \node [circle, draw=black , fill=black, inner sep=0pt, minimum size=2mm] (1) at (0,3) {};
    \node [circle, draw=black , fill=black, inner sep=0pt, minimum size=2mm] (2) at (3,0) {};
    \node [circle, draw=black , fill=black, inner sep=0pt, minimum size=2mm] (3) at (-3,0) {};
    
    \node  (4) [] at (0,-0.5) {$z_1$};
    \node  (5) [] at (0,3.5) {$z_2$};
    \node  (6) [] at (3,-0.5) {$z_3$};
    \node  (7) [] at (-3,-0.5) {$z_4$};
    
    \draw (2) to  [] (0) [-, thick];
    \draw (1) to  [] (2) [-, thick];
    \draw (3) to  [] (0) [-, thick];
    \draw (1) to  [] (3) [-, thick];
\end{tikzpicture}} 
  
\caption{\textit{Toric diagram of $\frac{\mathbb{C}^2}{\mathbb{Z}_2} \times \mathbb{C}$.}}
\label{fig23}
\end{figure}
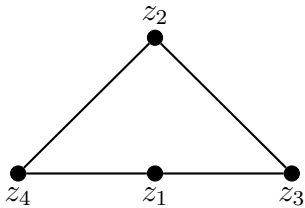\\
We have $m-n=4-3=1$ charge vector given by
\begin{equation}
    \begin{pmatrix}
0 & 0 & 1 & -1\\
0 & 1 & 0 & 0\\
1 & 1 & 1 & 1 \\
\end{pmatrix}\begin{pmatrix}
Q_1 \\
Q_2 \\
Q_3 \\
Q_4 \\
\end{pmatrix}=\Vec{0} \Rightarrow \begin{cases}
  Q_3-Q_4&=0 \\
  Q_2&=0 \\
   \sum_{i=1}^4 Q_i&=0 
\end{cases}
\end{equation}
and so $Q=(-2,0,1,1)$. Since we have one vanishing component, $Q_2=0$, the coordinate $z_2$ has no role and so we will expect a CY toric manifold of the form $X \times \mathbb{C}$, where $X$ is unknown for the moment and $\mathbb{C}$ is the space associated to $z_2$. We have the equivalence relation
\begin{equation}
    (z_1,z_2,z_3,z_4) \sim  (\lambda^{-2}z_1, z_2,\lambda z_3, \lambda z_4);
\end{equation}
since on $z_2$ the algebraic torus action is trivial we do not consider it and so we must find $m-1=4-1=3$ $\mathbb{C}^*$-invariant polynomials, for example
\begin{equation}
    Z_1=z_1z_3z_4, \ Z_2=z_1z^2_3, \ Z_3=z_1z_4^2,
\end{equation}
and they satisfy the relation
\begin{equation}
    Z_2Z_3=Z_1^2.
\label{rel3}    
\end{equation}
Relation (\ref{rel3}) is the realization of $\frac{\mathbb{C}^2}{\mathbb{Z}_2}$ as subvariety of $\mathbb{C}^3$. In the end, the toric CY variety is $\frac{\mathbb{C}^2}{\mathbb{Z}_2} \times \mathbb{C}$.\\

The general algorithm to identify to which variety a toric diagram belongs is the following: given the toric diagram we look at the equivalent relations that quotient out the algebraic torus action $(\mathbb{C}^*)^{m-n}$ and we construct a minimal basis of $m$ $(\mathbb{C}^*)^{m-n}$-invariant polynomials; the $m-n$ relations that these polynomials must satisfy identify the toric variety as subvariety of $\mathbb{C}^{m}$.

\subsubsection{Moment maps approach and Delzant-like construction}
Let us take a step back and consider a different way to define toric varieties. Let $\mathcal{M}$ be a symplectic manifold of real dimension $2n$ with symplectic form $\omega$. Given the action $U(1)^n \times \mathcal{M} \rightarrow \mathcal{M}$ this is said to be hamiltonian if its restriction to any $U(1) \subset U(1)^n$ is hamiltonian and any two of them commute. In this context the so-called moment map $\mu : \mathcal{M} \rightarrow \mathbb{R}^n$ emerges, whose components are the hamiltonians of each $U(1)$ action. Since all the $U(1)$ commute, for any $\Vec{r} \in \mathbb{R}^n$, its preimage $\mu^{-1}(\Vec{r})$ is invariant under the action of the full $U(1)^n$.  Moreover, if $\mathcal{M}$ has a Kähler structure, the existence of the hamiltonian action implies that the isometry group contains the algebraic torus  $U(1)^n \simeq (\mathbb{C}^*)^n$. So a toric variety emerges in a simple way as a real $2n$ dimensional symplectic manifold $\mathcal{M}$ that has an hamiltonian
action of the algebraic torus on it\footnote{There is a little caveat: if $\mathcal{M}$ is a cone the algebraic torus action must commute with the homothetic action induced by the Euler vector field.}. \\
We are interested in non-compact toric varieties but let us talk a little about compact ones. If $\mathcal{M}$ is compact, Delzant \cite{del} showed that the image through the moment map of the variety, $\mu(\mathcal{M})$, is a convex polytope $\Delta$ called Delzant polytope; however we are interested in CY cones, i.e. non-compact toric varieties. The generalization to a Deltzan-like construction is possible \cite{lerman2002contact} but the image under the moment map 
it is no longer a polytope but a cone:
\begin{equation}
    \Theta=\{\Vec{r} \in \mathbb{R}^n |\Vec{r} \cdot \Vec{v}_i \leq 0, \Vec{v}_i \in \mathbb{Z}^n\}
\end{equation}
and its dual graph is still a fan generated by the normal vectors $\Vec{v}_i$. Calabi-Yau condition imposes that $\Vec{v}_i$ are coplanar so we can project out the common component and get an $(n-1)$-dimensional object that encode the geometry: the toric diagram. Since the components of the vectors $\Vec{v}_i$ are integers, the toric diagram is the convex hull of a set of point in a $\mathbb{Z}^{n-1}$-lattice. In Deltzan-like approach, toric variety is built up using Kähler quotient and the toric variety is a $U(1)^n$ fibration over $\Theta$: one $U(1) \subset U(1)^n$ shrinks on the edges and so acts trivially, moreover, since in a vertex $n$ edges meet each others the full $U(1)^n$ fiber shrinks. Hence the vertexes are the fixed loci of the full algebraic torus action. It is interesting to note that this construction is well known by physicists: this is the moduli space of the Gauged Linear Sigma Model (GLSM), it is a SUSY gauge theory with abelian gauge group $U(1)^{m-n}$ and $m$ chiral superfields $Z_1,...,Z_m$ with charges $Q^1,...,Q^{m-n}$ under the $m-n$ $U(1)$.

In the case of our interest, namely CY threefold, the toric diagram is a two dimensional object and the CY cone threefold exhibits a three dimensional algebraic torus action. From the toric diagram we can get information on the dual QFT using the five brane system and brane tiling constructions; moreover we can calculate some useful quantities of the dual field theory directly from the toric diagram, such as the central charge, using the Butti and Zaffaroni procedure \cite{2005butti} or the symplexic decomposition procedure \cite{Manzoni:2022gxf}.
\subsection{Orientifold projections}\label{2a}
 Denoting by $0 \leq \sigma \leq \pi$ the coordinate describing the open string at a given time, the two ends $\sigma=0,\pi$ contain, thanks to Chan-Paton indexes, the gauge group degrees of freedom and the corresponding charged matter fields. At the endpoints we can apply Dirichlet or Neumann boundary conditions and we know that under $T$-duality these are interchanged. Orientifold projections of Type IIB theory are obtained by projecting the Type IIB spectrum by the involution $\Omega$ ($\Omega^2=1$), exchanging the left and right closed oscillators $\hat{\alpha}_m^{\mu}$, $\hat{\beta}_m^{\mu}$ and acting on the open strings as phases:
\begin{equation}
\begin{aligned}
 &closed \ strings \Rightarrow \Omega X^{\mu}(\tau,\sigma)\Omega^{-1}=X^{\mu}(\tau,-\sigma) \Rightarrow \hat{\alpha}_m^{\mu} \leftrightarrow \hat{\beta}_m^{\mu},\\
 &open \ strings \Rightarrow \Omega X^{\mu}(\tau,\sigma)\Omega^{-1}=X^{\mu}(\tau,\pi-\sigma) \Rightarrow \hat{\alpha}_m^{\mu} \rightarrow \pm(-1)^m\hat{\alpha}_m^{\mu}.
\end{aligned}    
\end{equation}
The overall effect is that orientifold projections map oriented strings to unoriented ones and type I superstring theory is nothing but an orientifold projection of type IIB superstring theory \cite{ANGELANTONJ_2002}: $\mathrm{type \ I}=\frac{\mathrm{type \ IIB}}{\Omega}$.

Orientifold introduces, from the worldsheet point of view, new surfaces in the Polyakov topological expansion. Indeed, due to the presence of unoriented strings, also non-orientable surfaces such as Klein bottle or Möbius strip are allowed. From the space-time viewpoint, these correspond to not dynamical objects, called O-planes, defined by $T$-duality as fixed point of the orientifold projection \cite{Dai:1989ua}. 

However this picture is not the complete one since, nowadays, there are no results in actual string theory for what one might expect to be the curved back-reacted geometry of orientifold. Indeed as reported in \cite{Cordova:2019cvf}, for any solution with O-planes, the presence of such a plane is inferred by comparison with their flat-space behavior. However, since the orientifolded space-times have strong curvature and coupling, stringy corrections come into play, and it is impossible to decide with supergravity alone whether the solutions are valid. Therefore, it is natural to think of more general behaviors in which geometric phase transitions due to the presence of orientifold can play a crucial role. Since this kind of behaviors are highly non-perturbative we expect something happens in the field theory that goes beyond the large $N$ limit.

This orientifold construction has important consequences on the emergent field theory build up with the AdS/CFT and toric CY machinery: some degrees of freedom are projected out and orthogonal and symplectic groups are now allowed together with symmetric and antisymmetric representations of unitary groups. Rules for construction of these kind of theories are studied for example in \cite{2007do,2014uno,Argurio:2020dko,Argurio:2020npm}. The crucial point here is that orientifold can give rise to different scenarios: 
\begin{itemize}
    \item in scenario I there is a new superconformal fixed point and the $R$-charges of the operators that are not projected out by $\Omega$ are the same as the charges of the corresponding one in the oriented theory and in the large $N$ limit. In the end, the post-orientifold central charge $a^{\Omega}$ turns out to be half of the pre-orientifold central charge $a$;
    \item scenario II does not admit a new superconformal RG fixed point;
    \item in scenario III there is a new superconformal fixed point but the $R$-charges of the operators that are not projected out by $\Omega$ are different from the charges of the corresponding one in the oriented theory and not only at large $N$. In the end, the post-orientifold central charge $a^{\Omega}$ turns out to be less than half of the pre-orientifold central charge $a$. However, something very interesting happens here: this could be (one of) the right field theory dual to the non-perturbative action of orientifold in the string theory side.
\end{itemize}
\subsubsection{The scenario $\mathrm{I}$ orientifold}
The scenario I orientifold occurs when the pre and post orientifold $R$-charges of the theories are the same and the central charge post orientifold is half of that pre orientifold:
\begin{equation}
    a^{\Omega}_i=a_i, a^{\Omega}=\frac{a}{2}.
\end{equation}
The Calabi-Yau cone describing the theory before and after the orientifold is then the same but with different volume due to the orientifold action. Moreover, due the $c$-theorem in $4D$, according to which $a_{IR}<a_{UV}$, scenario I orientifolds lead to the IR regimes of the theory.

\subsubsection{The  scenario $\mathrm{II}$ orientifold}
The scenario II seems to be quite trivial: there is no new superconformal point and the orientifold breaks conformal symmetry. However, as known in literature, these situation can give rise to a duality cascade or conformal symmetry can be restored with flavor branes \cite{Argurio:2017upa}.

\subsubsection{The scenario $\mathrm{III}$ orientifold and the IR duality}
In the scenario III orientifold the pre and post orientifold $R$-charges of theory are not the same and the post orientifold central charge is less than half of that pre orientifold \cite{2020in},\cite{2022su}:
\begin{equation}
     a^{\Omega}_i \neq a_i \ , \ a^{\Omega}<\frac{a}{2}.
\end{equation}
In this case the Calabi-Yau cones describing the theory before and after the orientifold are not the same. Scenario III orientifold seems to be part of a bigger picture where geometry/topology transitions play a crucial role. 
Note that, since the $c$-theorem in $4D$ tells us that $a_{IR}<a_{UV}$, also scenario III orientifolds lead to the IR regimes of the theory. 

\section{Algebro-geometrical orientifold and IR dualities}\label{3}
The IR duality we are talking about was recently proposed in \cite{2020in} studying the orientifold projections of Pseudo del Pezzo models (PdP) and subsequently extended to an infinite class of models in \cite{2022su}. Given two different oriented models $o\mathrm{CFT_A}$ and $o\mathrm{CFT_B}$, which have nothing to do one with each other, turns out that exist orientifold projections such that the $R$-charges, central charges and superconformal indexes of the two unoriented $u\mathrm{CFT_A}$ and $u\mathrm{CFT_B}$ models are the same \cite{2020in,2022su}
\begin{equation}
    a_i^{\Omega}=b_i^{\Omega}=b_i \ , \
    a_A^{\Omega}=a_B^{\Omega}=\frac{a_B}{2} \ , \ \mathcal{I}_A^{\Omega}=\mathcal{I}_B^{\Omega}.
\label{pino}
\end{equation}
This suggests the proposed duality and since orientifold leads to the IR regime, this is an IR duality. More specifically the $o\mathrm{CFT_A}$ has a scenario I orientifold while the $o\mathrm{CFT_B}$ has a scenario III orientifold and the set of $R$-charges, central charges and superconformal indexes of the two unoriented models are the same for any finite value of $N$. 
What we have said in words is summarized in the following scheme. 
\begin{equation}
\begin{aligned}
&(a_i,a_{\mathrm{A}}) \ \ \ \ \ \ \ \ \ \ \ \ \ \ \ \ \ \ \ \ \ \ \ \ \ \ \ \ \ \ \ \ \ \ \ \ \ \ \ \ \ \ \ \ \ \ \ \  \ \ \ \ \left(a^{\Omega}_i\neq a_i, a_{\mathrm{A}}^{\Omega}<\frac{a_A}{2}\right)\\
&oriented \ model \ \mathrm{A} \xrightarrow{scenario \ \mathrm{III} \ orientifold \ projection} unoriented \ model \ \mathrm{A}\\
&\ \ \ \ \ \ \ \ \ \ \ \ \ \ \ \ \ \ \ \ \ \ \ \ \ \ \ \ \ \ \ \ \ \ \ \ \ \ \ \ \ \ \ \ \ \ \ \  \ \ \ \ \ \ \ \ \ \ \ \ \ \ \ \ \mathrm{IR} \ duals  \Bigg\updownarrow\\
&oriented \ model \ \mathrm{B} \xrightarrow{scenario \ \mathrm{I} \ orientifold \ projection \ } unoriented \ model \ \mathrm{B}\\
&(b_i,a_{\mathrm{B}}) \ \ \ \ \ \ \ \ \ \ \ \ \ \ \ \ \ \ \ \ \ \ \ \ \ \ \ \ \ \ \ \ \ \ \ \ \ \ \ \ \ \ \ \ \ \ \ \  \ \ \ \ \left(b^{\Omega}_i=b_i, a_{\mathrm{B}}^{\Omega}=\frac{a_{\mathrm{B}}}{2}\right)\\
\end{aligned}    
\end{equation}
Some examples of the IR duality due to orientifold in scenario III seems to be explained and interpreted, from the field theory point of view, in terms of inherited S-duality from the $\mathcal{N}=2$ case. The authors observe that the duality for the $\mathcal{N}=1$ models discussed in \cite{2022su} corresponds to S-duality at different points of the conformal manifold \cite{Amariti_2022,Amariti:2022dyi,Amariti:2022dui}. Therefore the duality between the two unoriented models can be thought as a conformal duality. The crucial property behind this result is the presence in the spectrum of two-index tensor fields with R-charge equal to one, uncharged with respect to the other global symmetries. 

Our main goal is to better understand the link between orientifolds and the IR duality from the point of view of the geometry. The guiding ideas are the flop transition and the other geometry/topology changing in string theory compactifications, where the geometry/topology of CY manifold are modified due to a quantum behavior of the geometry itself \cite{Greene:1996cy,Aspinwall:1993nu,Greene:1995hu,Witten:1993yc}. Heuristically, one suspects that geometry/topology might be able to change by means of the violent curvature fluctuations, such as induced by orientifold, which would be expected in any quantum theory of gravity. Therefore, the basic idea is that orientifold projections map the algebraic equations describing the geometry of the CY$_{\mathrm{A}}$ into those that describe the geometry of the CY$_{\mathrm{B}}$. This orientifold modifies the geometry of the CY$_{\mathrm{A}}$ cone due to its intrinsic non-perturbative nature and this geometrical change induces $k$ new degrees of freedom with $R$-charges $\tilde{a}_i$ to emerge (we will call them orientifold-mapping operators or $o\mu $-operators). They make the original field theory no longer fulfilling the condition $\sum_1^d a_i+\sum_1^k \tilde{a}_i=2$; hence old $R$-charges $a_i$ and $o\mu$-operators $R$-charges $\tilde{a}_i$ must mix together to return a set of new $R$-charges $a^{\Omega}_i=b_i$ consistent with the modified CY$^{\Omega}_{\mathrm{A}}$ cone geometry, that is now CY$_\mathrm{B}$ cone, and satisfying the condition $\sum_1^d b_i=2$. Steps are summarized in the following scheme.

\begin{equation}
\begin{aligned}
&(a_i,\mathrm{CY_{{A}}}) \ \ \ \ \ \ \ \ \ \ \ \ \ \ \ \ \ \ \ \ \ \ \ \ \ \ \ \ \ \ \ \ \ \ \ \ \ \ \ \ \ \ \ \ \ \ \ \  \ \ \ \ \ \ \  \left(a^{\Omega}_i\neq a_i, \tilde{a}_i, \mathrm{CY}^{\Omega}_{\mathrm{A}} = \mathrm{CY_B} \right)\\
&oriented \ model \ \mathrm{A} \ \ \ \ \xrightarrow{scenario \ \mathrm{I}\ orientifold \ projection} \ \ \ unoriented \ model \ \mathrm{A}\\
&\ \ \ \ \ \ \ \ \ \ \ \ \ \ \ \ \ \ \ \ \ \ \ \ \ \ \ \ \ \ \ \ \ \ \ \ \ \ \ \ \ \ \ \ \ \ \ \  \ \ \ \ \ \ \ \ \ \mathrm{IR} \ duals  \Bigg\downarrow (a_i \ \mathrm{and} \ \tilde{a}_i \ R\mathrm{-charges \ mixing}) \\
&oriented \ model \ \mathrm{B} \ \ \ \ \xrightarrow{scenario \ \mathrm{I} \ orientifold \ projection \ } \ \ \ unoriented \ model \ \mathrm{B}\\
&(b_i,\mathrm{CY_{{B}}}) \ \ \ \ \ \ \ \ \ \ \ \ \ \ \ \ \ \ \ \ \ \ \ \ \ \ \ \ \ \ \ \ \ \ \ \ \ \ \ \ \ \ \ \ \ \ \ \  \ \ \ \ \ \  \ \ \left(b^{\Omega}_i=b_i,\mathrm{CY_{{B}}} \right)\\
\label{schema}
\end{aligned}    
\end{equation}
The procedure is the following: from toric diagrams of theories A and B we determine the sets of algebraic equations using toric geometry tools exposed in Section \ref{2}; then we map these two sets into each other and, finally, we associate at every homogeneous coordinate $z_i$ a trial $R$-charge, adding the minimal number of new degrees of freedom. 
The analysis must consider two different cases:
\begin{enumerate}
    \item $\Delta P_e:=P^{(\mathrm{B})}_e-P_e^{(\mathrm{A})}=0$;
    \item $\Delta P_e:=P^{(\mathrm{B})}_e-P_e^{(\mathrm{A})}=1$;
\end{enumerate}
$P^{(\bullet)}_e$ is the number of extremal point of the toric diagram of the theory $\bullet$. Since in the oriented theory only the trial $R$-chagres associated to extremal points matter, in the first case the two oriented theories have the same number of $R$-charges while in the second case the number of $R$-charges of the oriented theories are not equal and we expect that some not extremal points of the toric diagram of theory $\mathrm{A}$ became extremal points after the orientifold action. Cases with $\Delta P_e>1$ are not discussed in this work since no examples have been found; they may be studied in future works to better understand the link between geometry, orientifold and the IR duality. 

The interpretation of orientifold in this way will be called an algebro-geometrical orientifold. This orientifold is different from scenario I and III orientifolds because, referring to scheme (\ref{schema}), the latters act horizontally while the former acts crosswise from the oriented model $\mathrm{A}$ to the unoriented model $\mathrm{B}$. This kind of orientifold implements the IR duality from the geometric point of view as morphism between Calabi-Yau algebraic varieties. The schematic picture to have in mind is Figure \ref{fig7}.

In following paragraph we study these two cases with specific examples but this construction can be repeated for many other pairs and the result is that it is always possible to construct relevant operators with the new degrees of freedom inserted by matching the systems containing the polynomial equations that define the geometries of the pair.

\subsection{Different number of external points: $\Delta P_e=1$}
Let us consider the conifold $\mathcal{C}$ and  $\frac{\mathbb{C}^2}{\mathbb{Z}_2}\times \mathbb{C}$; for toric diagrams and toric data we refer to Examples in Paragraph \ref{exc3}. Moreover, in those examples determination of the equations that describe the CY cones have already been done. The conifold has scenario I orientifold while $\frac{\mathbb{C}^3}{\mathbb{Z}_2}$ has scenario III orientifold and the two theories are IR duals. 
The set of equations for the two theories are 
\begin{equation}
\begin{aligned}
& \frac{\mathbb{C}^2}{\mathbb{Z}_2}\times \mathbb{C}: \ \ \ \ \ \ \ \ \ \ \ \ \ \ \ \ \ \ \ \ \ \ \ \ \ \ \ \ \ \ \ \ \  \mathcal{C}: \\
&Y_2Y_3=Y_1^2; \ \ \ \ \ \ \ \ \ \ \ \ \ \ \ \ \ \ \ \ \ \ \ X_1X_4=X_2X_3; \\
&\begin{cases}
  Y_1&=z_1z_3z_4; \\
  Y_2&=z_1z_3^2; \\
  Y_3&=z_1z_4^2. 
\end{cases} \ \ \ \ \ \ \ \ \ \ \ \ \ \ \ \begin{cases}
  X_1&=z_1z_2; \\
  X_2&=z_1z_4; \\
  X_3&=z_2z_3; \\
  X_4&=z_3z_4.
\end{cases}
\end{aligned}
\end{equation}
Now we have to map the two equations in all possible ways remembering that after the orientifold action the toric diagrams are mapped into each other and so the coordinates $z_i$ are effectively the same. However, if we want to keep track of the different $R$-charges associated to the homogeneous coordinate $z_i$ of the two different toric diagrams, so the $R$-charges $b_i=b_i^{\Omega}$ and $a_i$, we must remember that we have to associate a different set of $R$-charges at those $z_i$ belonging to different equations and so different theories. We have to consider the following mappings:
\begin{equation}
\begin{aligned}
\mathrm{I}:&\begin{cases}
  Y^2_1&=X_2X_3; \\
  Y_2&=X_1; \\
  Y_3&=X_4;
\end{cases} \ \ \ \ \ \ \ \ \ \ \ \ \ \ \ \ \mathrm{II}: \ \begin{cases}
  Y^2_1&=X_2X_3; \\
  Y_2&=X_4; \\
  Y_3&=X_1;
\end{cases}\\
\mathrm{III}:& \begin{cases}
  Y^2_1&=X_1X_4; \\
  Y_2&=X_2; \\
  Y_3&=X_3;
\end{cases} \ \ \ \ \ \ \ \ \ \ \ \ \ \ \ \ \mathrm{IV}: \ \begin{cases}
  Y^2_1&=X_1X_4; \\
  Y_2&=X_3; \\
  Y_3&=X_2;
\end{cases}
\end{aligned}    
\end{equation}
By associating the set of oriented $R$-charges $\{z_1,z_2,z_3,z_4\} \rightarrow \{0,a_2,a_3,a_4\}$ to the $z_i$ coordinates of $Y_i$ polynomials and the set of oriented $R$-charges $\{z_1,z_2,z_3,z_4\} \rightarrow \{b_1,b_2,b_3,b_4\}$ to the $z_i$ coordinates of $X_i$ polynomials, we get the following relations:
\begin{equation}
\begin{aligned}
\mathrm{I}:&\begin{cases}
  2a_3+2a_4&=b_1+b_2+b_3+b_4; \\
  2a_3&=b_1+b_2; \\
  2a_4&=b_3+b_4;
\end{cases} \ \ \ \ \ \ \ \ \mathrm{II}: \ \begin{cases}
  2a_3+2a_4&=b_1+b_2+b_3+b_4; \\
  2a_3&=b_3+b_4; \\
  2a_4&=b_1+b_2;
\end{cases}\\
\mathrm{III}:& \begin{cases}
  2a_3+2a_4&=b_1+b_2+b_3+b_4; \\
  2a_3&=b_1+b_4; \\
  2a_4&=b_2+b_3;
\end{cases} \ \ \ \ \ \ \ \ \mathrm{IV}: \ \begin{cases}
  2a_3+2a_4&=b_1+b_2+b_3+b_4; \\
  2a_3&=b_2+b_3; \\
  2a_4&=b_1+b_4;
\end{cases}
\end{aligned}      
\label{ugo}
\end{equation}
and they are not all independent. 
If we make together systems I, II, III and IV we obtain the following relations:
\begin{equation}
\begin{cases}
 \mathrm{I}=\mathrm{III} \Rightarrow b_1+b_2=b_1+b_4 &\Rightarrow b_2=b_4; \\
  \mathrm{I}=\mathrm{IV} \Rightarrow b_3+b_4=b_1+b_2  &\Rightarrow b_1=b_3;
\end{cases}
\end{equation}
and we have also the relation $\sum_{i=1}^4b_i=2$. These conditions are not sufficient to fix unequivocally the values of the $R$-charges to those of the conifold. However this is right, indeed using the fact we know the values at the superconformal fixed point of the two set of $R$-charges, namely $\{a_1,a_2,a_3,a_4\}=\{0,2/3,2/3,2/3\}$ and  $\{b_1,b_2,b_3,b_4\}=\{1/2,1/2,1/2,1/2\}$, we may note that the equations (\ref{ugo}) do not do the right job. Nevertheless, if we admit the existence of a new operator (an orientifold-mapping operator) with $R$-charge $\tilde{a}=\frac{1}{3}$ we note that, for example,
\begin{equation}\mathrm{I}:\begin{cases}
  2a_3+2a_4-2\tilde{a}&=b_1+b_2+b_3+b_4; \\
  2a_3-\tilde{a}&=b_1+b_2; \\
  2a_4-\tilde{a}&=b_3+b_4;
\end{cases} \Rightarrow 
 \begin{cases}
  2\frac{2}{3}+2\frac{2}{3}-2\frac{1}{3}&=\frac{1}{2}+\frac{1}{2}+\frac{1}{2}+\frac{1}{2}; \\
  2\frac{2}{3}-\frac{1}{3}&=\frac{1}{2}+\frac{1}{2}; \\
  2\frac{2}{3}-\frac{1}{3}&=\frac{1}{2}+\frac{1}{2};
\end{cases}   
\end{equation}
now we have the right matching. 

Finally, let us now consider the fields associated to $\frac{\mathbb{C}^2}{\mathbb{Z}_2} \times \mathbb{C}$ and their $R$-charges \cite{2005butti,Manzoni:2022gxf}.
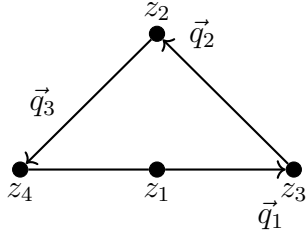
\begin{figure}[ht]
\centering{\begin{tikzpicture}[ auto, scale=0.6] 
    \node [circle, draw=black , fill=black, inner sep=0pt, minimum size=2mm] (0) at (0,0) {};
    \node [circle, draw=black , fill=black, inner sep=0pt, minimum size=2mm] (1) at (0,3) {};
    \node [circle, draw=black , fill=black, inner sep=0pt, minimum size=2mm] (2) at (3,0) {};
    \node [circle, draw=black , fill=black, inner sep=0pt, minimum size=2mm] (3) at (-3,0) {};
    
    \node  (4) [] at (0,-0.5) {$z_1$};
    \node  (5) [] at (0,3.5) {$z_2$};
    \node  (6) [] at (3,-0.5) {$z_3$};
    \node  (7) [] at (-3,-0.5) {$z_4$};
    
    \draw (2) to  [] (0) [<-, thick];
    \draw (1) to  [] (2) [<-, thick];
    \draw (3) to  [] (0) [-, thick];
    \draw (1) to  [] (3) [->, thick];

    \node  (8) [] at (-2.5,1.5) {$\Vec{q}_3$};
    \node  (9) [] at (1,3) {$\Vec{q}_2$};
    \node  (10) [] at (2.5,-1) {$\Vec{q}_1$};
\end{tikzpicture}} 
  
\caption{\textit{Toric diagram of $\frac{\mathbb{C}^2}{\mathbb{Z}_2} \times \mathbb{C}$ with the useful vectors to associate fields and $R$-charges.}}
\label{toricharge1}
\end{figure}\\
Referring to Figure \ref{toricharge1} we have
\begin{equation}
\begin{aligned}
&\langle \Vec{q}_1,\Vec{q}_2 \rangle=2 \rightarrow a_3\\
&\langle \Vec{q}_2,\Vec{q}_3 \rangle=2 \rightarrow a_2\\
&\langle \Vec{q}_3,\Vec{q}_1 \rangle=2 \rightarrow a_4, a_4+a_1
\end{aligned}    
\end{equation}
where 
\begin{equation*}
\langle \Vec{u},\Vec{v} \rangle:=det \begin{bmatrix}
u^{(1)} & u^{(2)} \\
v^{(1)} & v^{(2)}
\end{bmatrix}
\; . \label{eq:uvdet}
\end{equation*}
Hence we have 2 fields with $R$-charge $a_3$, 2 fields with $R$-charge $a_2$, 1 field with $R$-charge $a_4$ and 1 field with $R$-charge $a_4+a_1$. Despite in the oriented theory charge $a_1$ does not play any role, here, in the unoriented one, we can identify $a_1\equiv \tilde{a}$ and so there is a field with a new $R$-charge due to orietifold action; we call it $\Pi$\footnote{Its bosonic part will be dubbed puppon while its fermionic one puchino. This field is the one associated, according to \cite{2005butti} or \cite{Manzoni:2022gxf}, to the $R$-charge given by the combination of the new charge entered in the game, $\tilde{a}$, and the other charges.}. It is interesting to note that field $\Pi$ associated with the $R$-charge $a_4+\tilde{a}$ has $R=a_4+\tilde{a}=\frac{2}{3}+\frac{1}{3}=1$. We can build up a relevant deformation with $\Pi$, for example $\mathcal{O}_i=\Phi_i \Pi$ where $\Phi_i$ are the other chiral fields with $R$-charge $R[\Phi_i]=\frac{2}{3}$. The scaling dimension of $\mathcal{O}_i $ is given by $\Delta_{\mathcal{O}_i }=\frac{3}{2}R[\mathcal{O}_i]=\frac{5}{2}<3$. This makes sense given that orientifold has the effect to make the theory flow
into the IR regime; so this operator turns on and the theory becomes IR dual to another theory that has the correct set of $R$-charges to match systems (\ref{ugo}). \\
One may wonder since the mapping is not one to one; however this seems to be a peculiarity of very symmetric models described by highly symmetric toric diagrams with small number of points. Indeed in the case of $\mathcal{C}$ and $\frac{\mathbb{C}^2}{\mathbb{Z}_2}\times \mathbb{C}$ there are exactly four ways of superimposing the two toric diagrams so as to place the greatest number of points of the two diagrams in the same position. From the field theory point of view this ambiguity is probably due to the flavour global symmetries of the models since in both the set of $R$-charges the $R$-charges itself are identical.

\subsection{Equal number of external points: $\Delta P_e=0$}
Consider the two theories $\frac{\mathrm{SPP}}{\mathbb{Z}_2}$ and $L^{(3,3,3)}$; their toric diagrams are drawn below
\begin{figure}[h]
\centering{\begin{tikzpicture}[ auto, scale=0.6] 
    \node [circle, draw=black , fill=black, inner sep=0pt, minimum size=2mm] (0) at (0,0) {};
    \node [circle, draw=black , fill=black, inner sep=0pt, minimum size=2mm] (1) at (1,0) {};
    \node [circle, draw=black , fill=black, inner sep=0pt, minimum size=2mm] (2) at (1,1) {};
    \node [circle, draw=black , fill=black, inner sep=0pt, minimum size=2mm] (3) at (1,2) {};
    \node [circle, draw=black , fill=black, inner sep=0pt, minimum size=2mm] (4) at (1,3) {};
    \node [circle, draw=black , fill=black, inner sep=0pt, minimum size=2mm] (5) at (0,3) {};
    \node [circle, draw=black , fill=black, inner sep=0pt, minimum size=2mm] (6) at (0,2) {};
    \node [circle, draw=black , fill=black, inner sep=0pt, minimum size=2mm] (7) at (0,1) {};
    
    \node  (8) [] at (1.5,3.5) {$z_1$};
    \node  (9) [] at (-0.5,3.5) {$z_2$};
    \node  (10) [] at (-0.5,2) {$z_3$};
    \node  (11) [] at (-0.5,1) {$z_4$};
    \node  (12) [] at (-0.5,-0.5) {$z_5$};
    \node  (13) [] at (1.5,-0.5) {$z_6$};
    \node  (14) [] at (1.5,1) {$z_7$};
    \node  (15) [] at (1.5,2) {$z_8$};
    
    \draw (0) to  [] (1) [-, thick];
    \draw (1) to  [] (2) [-, thick];
    \draw (2) to  [] (3) [-, thick];
    \draw (3) to  [] (4) [-, thick];
    \draw (4) to  [] (5) [-, thick];
    \draw (5) to  [] (6) [-, thick];
    \draw (6) to  [] (7) [-, thick];
    \draw (7) to  [] (0) [-, thick];
\end{tikzpicture}} 
\centering{\begin{tikzpicture}[ auto, scale=0.6] 
    \node [] (0) at (0,0) {};
    \node [] (0) at (1,0) {};
\end{tikzpicture}} 
\centering{\begin{tikzpicture}[ auto, scale=0.6] 
    \node [circle, draw=black , fill=black, inner sep=0pt, minimum size=2mm] (0) at (1,-1) {};
    \node [circle, draw=black , fill=black, inner sep=0pt, minimum size=2mm] (1) at (1,0) {};
    \node [circle, draw=black , fill=black, inner sep=0pt, minimum size=2mm] (2) at (1,1) {};
    \node [circle, draw=black , fill=black, inner sep=0pt, minimum size=2mm] (3) at (1,2) {};
    \node [circle, draw=black , fill=black, inner sep=0pt, minimum size=2mm] (4) at (1,3) {};
    \node [circle, draw=black , fill=black, inner sep=0pt, minimum size=2mm] (5) at (0,3) {};
    \node [circle, draw=black , fill=black, inner sep=0pt, minimum size=2mm] (6) at (0,2) {};
    \node [circle, draw=black , fill=black, inner sep=0pt, minimum size=2mm] (7) at (0,1) {};
    
    \node  (8) [] at (1.5,3.5) {$z_1$};
    \node  (9) [] at (-0.5,3.5) {$z_2$};
    \node  (10) [] at (-0.5,2) {$z_3$};
    \node  (11) [] at (-0.5,0.5) {$z_4$};
    \node  (12) [] at (1.5,-1.5) {$z_5$};
    \node  (13) [] at (1.5,0) {$z_6$};
    \node  (14) [] at (1.5,1) {$z_7$};
    \node  (15) [] at (1.5,2) {$z_8$};
    
    \draw (0) to  [] (1) [-, thick];
    \draw (1) to  [] (2) [-, thick];
    \draw (2) to  [] (3) [-, thick];
    \draw (3) to  [] (4) [-, thick];
    \draw (4) to  [] (5) [-, thick];
    \draw (5) to  [] (6) [-, thick];
    \draw (6) to  [] (7) [-, thick];
    \draw (7) to  [] (0) [-, thick];
\end{tikzpicture}} 
\caption{\textit{Toric diagrams of $\frac{\mathrm{SPP}}{\mathbb{Z}_2}$ (right) and $L^{(3,3,3)}$ (left)}.}
\label{fig23}
\end{figure}
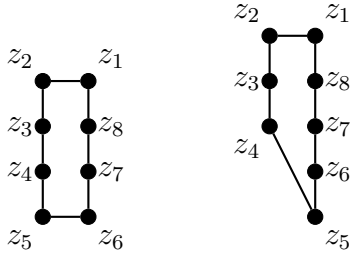\\
To determine the set of equations describing the two varieties we follow the step highlighted in Section \ref{2}. \\
Let us start with $\frac{\mathrm{SPP}}{\mathbb{Z}_2}$. We have $m-n=8-3=5$ charge vectors solution of
\begin{equation}
    \begin{pmatrix}
1 & 0 & 0 & 0 & 1 & 1 & 1 & 1\\
3 & 3 & 2 & 1 & -1 & 0 & 1 & 2\\
1 & 1 & 1 & 1 & 1 & 1 & 1 & 1 \\
\end{pmatrix}\begin{pmatrix}
Q_1^a \\
Q_2^a \\
Q_3^a \\
Q_4^a \\
Q_5^a \\
Q_6^a \\
Q_7^a \\
Q_8^a \\
\end{pmatrix}=\Vec{0} \Rightarrow \begin{cases}
  Q^a_1+Q^a_5+Q^a_6+Q^a_7+Q^a_8&=0 \\
  3Q_1^a+3Q_2^a+2Q_3^a+Q_4^a-Q_5^a+Q_7^a+2Q_8^a&=0 \\
   \sum_{i=1}^8 Q^a_i&=0 
\end{cases};
\end{equation}
where $a=1,...,5$. Solution space can be parameterized as follows:
\begin{equation}
\begin{cases}
&Q_1^a=-Q^a_5-Q^a_6-Q^a_7-Q^a_8;\\
&Q_2^a=Q^a_4+4Q^a_5+3Q^a_6+2Q^a_7+Q^a_8;\\
&Q_3^a=-2Q_4^a-4Q_5^a-3Q_6^a-2Q_7^a-Q_8^a;
\label{gino}
\end{cases}
\end{equation}
the equivalence relations are given by
\begin{equation}
\begin{aligned}
    &(z_1,z_2,z_3,z_4,z_5,z_6,z_7,z_8) \sim  \\
    &\sim (\lambda^{Q_1^a}z_1,\lambda^{Q_2^a}z_2,\lambda^{Q_3^a}z_3,\lambda^{Q_4^a}z_4,\lambda^{Q_5^a}z_5,\lambda^{Q_6^a}z_6,\lambda^{Q_7^a}z_7,\lambda^{Q_8^a}z_8),
\end{aligned}    
\end{equation}
where $Q_1^a,Q_2^a,Q_3^a$ satisfy (\ref{gino}). The 8 invariant polynomials can be constructed in a simply way: as first step we consider the homogeneous coordinates with not independent charges (in this case $z_1,z_2,z_3$) and then we multiply them by the right combination of the other $z_i$, in order to compensate the transformation under the equivalence relations; the second step is to build up the other polynomials taking product or quotient of polynomials constructed at the first step. This procedure not only gives us the invariant polynomials but also the relations between them.
Invariant polynomials and relations are given by
\begin{equation}
\begin{cases}
&X_1=z_1z_5z_6z_7z_8;\\
&X_2=z_2z_4^{-1}z_5^{-4}z_6^{-3}z_7^{-2}z_8^{-1};\\
&X_3=z_3z_4^2z_5^4z_6^3z_7^2z_8;\\
&X_4=z_1z_2z_4^{-1}z_5^{-3}z_6^{-2}z_7^{-1}=X_1X_2;\\
&X_5=z_1z_3z_4^2z_5^5z_6^4z_7^3z_8^2=X_1X_3;\\
&X_6=z_2z_3z_4=X_2X_3;\\
&X_7=z_1^{-1}z_3z_4^2z_5^3z_6^2z_7=\frac{X_3}{X_1};\\
&X_8=z_2^{-1}z_3z_4^3z_5^8z_7^4z_8^2=\frac{X_3}{X_2}.
\label{setS}
\end{cases}
\end{equation}

For $L^{(3,3,3)}$ we follow the same steps. We have $m-n=8-3=5$ charge vectors solution of
\begin{equation}
    \begin{pmatrix}
1 & 0 & 0 & 0 & 0 & 1 & 1 & 1\\
3 & 3 & 2 & 1 & 0 & 0 & 1 & 2\\
1 & 1 & 1 & 1 & 1 & 1 & 1 & 1 \\
\end{pmatrix}\begin{pmatrix}
Q_1^a \\
Q_2^a \\
Q_3^a \\
Q_4^a \\
Q_5^a \\
Q_6^a \\
Q_7^a \\
Q_8^a \\
\end{pmatrix}=\Vec{0} \Rightarrow \begin{cases}
  Q^a_1+Q^a_6+Q^a_7+Q^a_8&=0 \\
  3Q_1^a+3Q_2^a+2Q_3^a+Q_4^a+Q_7^a+2Q_8^a&=0 \\
   \sum_{i=1}^8 Q^a_i&=0 
\end{cases};
\end{equation}
where $a=1,...,5$. Solution space can be parameterized as follows:
\begin{equation}
\begin{cases}
&Q_1^a=-Q^a_6-Q^a_7-Q^a_8;\\
&Q_2^a=Q^a_4+2Q^a_5+3Q^a_6+2Q^a_7+Q^a_8;\\
&Q_3^a=-2Q_4^a-3Q_5^a-3Q_6^a-2Q_7^a-Q_8^a;
\label{gino1}
\end{cases}
\end{equation}
the equivalence relations are given by
\begin{equation}
\begin{aligned}
    &(z_1,z_2,z_3,z_4,z_5,z_6,z_7,z_8) \sim  \\
    &\sim (\lambda^{Q_1^a}z_1,\lambda^{Q_2^a}z_2,\lambda^{Q_3^a}z_3,\lambda^{Q_4^a}z_4,\lambda^{Q_5^a}z_5,\lambda^{Q_6^a}z_6,\lambda^{Q_7^a}z_7,\lambda^{Q_8^a}z_8),
\end{aligned}    
\end{equation}
where $Q_1^a,Q_2^a,Q_3^a$ satisfy (\ref{gino1}).
Invariant polynomials and relations are given by
\begin{equation}
\begin{cases}
&Y_1=z_1z_6z_7z_8;\\
&Y_2=z_2z_4^{-1}z_5^{-2}z_6^{-3}z_7^{-2}z_8^{-1};\\
&Y_3=z_3z_4^2z_5^3z_6^3z_7^2z_8;\\
&Y_4=z_1z_2z_4^{-1}z_5^{-2}z_6^{-2}z_7^{-1}=Y_1Y_2;\\
&Y_5=z_1z_3z_4^2z_5^3z_6^4z_7^3z_8^2=Y_1Y_3;\\
&Y_6=z_2z_3z_4z_5=Y_2Y_3;\\
&Y_7=z_1^{-1}z_3z_4^2z_5^3z_6^2z_7=\frac{Y_3}{Y_1};\\
&Y_8=z_2^{-1}z_3z_4^3z_5^5z_6^6z_7^4z_8^2=\frac{Y_3}{Y_2}.
\label{setL}
\end{cases}
\end{equation} 
Let us now proceed with the matching between sets (\ref{setS}) and (\ref{setL}). Since we have $\Delta P_e=0$ we can not consider all the not extremal $z_i$, indeed we expect that no new point enters in the game. We note that if we match the first three polynomials of the two sets we get automatically that all other polynomials match, so
\begin{equation}
\begin{cases}
& X_1=Y_1 \Rightarrow z_1z_5=z_1z_6 \Rightarrow a_3+a_4=b_1+b_4;\\
& X_2=Y_2 \Rightarrow z_2z_4^{-1}z_5^{-4}=z_2z_5^{-2}z_6^{-3} \Rightarrow a_2-a_1-4a_4=b_2-2b_3-3b_4;\\
& X_3=Y_3 \Rightarrow z_4^2z_5^4=z_5^3z_6^3 \Rightarrow 4a_4+2a_1=3b_3+3b_4.
\label{peppe23}
\end{cases}    
\end{equation}
If we now consider the set of non-vanishing $R$-charges of $\frac{\mathrm{SPP}}{\mathbb{Z}_2}$ ($\{a_1,a_2,a_3,a_4\}=\{1-\frac{1}{\sqrt{3}},1-\frac{1}{\sqrt{3}},\frac{1}{\sqrt{3}},\frac{1}{\sqrt{3}}\}$) and $L^{(3,3,3)}$ ($\{b_1,b_2,b_3,b_4\}=\{\frac{1}{2},\frac{1}{2},\frac{1}{2},\frac{1}{2}\}$) we note that system (\ref{peppe23}) does not match unless we suppose the existence of a new operator (an orientifold-mapping operator) with $R$-charge $\tilde{a}=\frac{2-\sqrt{3}}{\sqrt{3}}$.

Following the same steps of the case $\Delta P_e=1$ we can associate $R$-charges to the fields of theory $\frac{\mathrm{SPP}}{\mathbb{Z}_2}$.
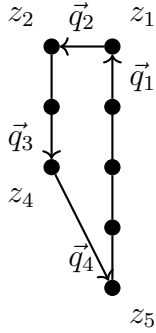
\begin{figure}[h]
\centering{\begin{tikzpicture}[ auto, scale=0.8] 
    \node [circle, draw=black , fill=black, inner sep=0pt, minimum size=2mm] (0) at (1,-1) {};
    \node [circle, draw=black , fill=black, inner sep=0pt, minimum size=2mm] (1) at (1,0) {};
    \node [circle, draw=black , fill=black, inner sep=0pt, minimum size=2mm] (2) at (1,1) {};
    \node [circle, draw=black , fill=black, inner sep=0pt, minimum size=2mm] (3) at (1,2) {};
    \node [circle, draw=black , fill=black, inner sep=0pt, minimum size=2mm] (4) at (1,3) {};
    \node [circle, draw=black , fill=black, inner sep=0pt, minimum size=2mm] (5) at (0,3) {};
    \node [circle, draw=black , fill=black, inner sep=0pt, minimum size=2mm] (6) at (0,2) {};
    \node [circle, draw=black , fill=black, inner sep=0pt, minimum size=2mm] (7) at (0,1) {};
    
    \node  (8) [] at (1.5,3.5) {$z_1$};
    \node  (9) [] at (-0.5,3.5) {$z_2$};
    \node  (11) [] at (-0.5,0.5) {$z_4$};
    \node  (12) [] at (1.5,-1.5) {$z_5$};
    
    \draw (0) to  [] (1) [-, thick];
    \draw (1) to  [] (2) [-, thick];
    \draw (2) to  [] (3) [-, thick];
    \draw (3) to  [] (4) [->, thick];
    \draw (4) to  [] (5) [->, thick];
    \draw (5) to  [] (6) [-, thick];
    \draw (6) to  [] (7) [->, thick];
    \draw (7) to  [] (0) [->, thick];

    \node  (16) [] at (1.5,2.5) {$\Vec{q}_1$};
    \node  (17) [] at (0.5,3.5) {$\Vec{q}_2$};
    \node  (18) [] at (-0.5,1.5) {$\Vec{q}_3$};
    \node  (19) [] at (0.5,-0.5) {$\Vec{q}_4$};
\end{tikzpicture}} 
\caption{\textit{Toric diagram of $\frac{\mathrm{SPP}}{\mathbb{Z}_2}$ with the useful vectors to associate fields and $R$-charges.}}
\label{toricharge2}
\end{figure}\\
Referring to Figure \ref{toricharge2} we have 
\begin{equation}
\begin{aligned}
&\langle \Vec{q}_1,\Vec{q}_2 \rangle = 4 \rightarrow a_3;\\
&\langle \Vec{q}_2,\Vec{q}_3 \rangle = 2 \rightarrow a_2;\\
&\langle \Vec{q}_2,\Vec{q}_4 \rangle = 2 \rightarrow a_1+a_2;\\
&\langle \Vec{q}_3,\Vec{q}_4 \rangle = 2 \rightarrow a_1 ;\\
&\langle \Vec{q}_4,\Vec{q}_1 \rangle = 4 \rightarrow a_4.
\end{aligned}
\end{equation}
In this case we do not have an obvious candidate to play the role of $\Pi$ since no new extremal point enters in the game. Despite this, we can construct relevant operators that perturb the conformal theory and induce the RG flow to the dual theory in the IR. As in the $\Delta P_e=1$ case we add the new $R$-charge to the point that must be moved to transform the toric diagram of $\frac{\mathrm{SPP}}{\mathbb{Z}_2}$ into the toric diagram of $L^{(3,3,3)}$, namely $z_5$. Therefore $\Pi$ is a field with $R$-charge $R[{\Pi}]=a_4+\Tilde{a}$ and a relevant operator can be constructed, for example, as $\mathcal{O}_i=\Pi \chi_i$ where $\chi_i$ are the two fields with $R$-charge $R[\chi_i]=a_1+a_2$. The scaling dimension of $\mathcal{O}_i$ is given by $\Delta[\mathcal{O}_i]=\frac{3}{2}R[\mathcal{O}_i]=\frac{3}{2}<3$.\\
In this case we have only one matching which is suggested by the fact that we have only one way of superimposing the two toric diagrams so as to place the greatest number of points of the two diagrams in the same position. Indeed the toric diagrams are less symmetric with respect to the case of $\mathcal{C}$ and $\frac{\mathbb{C}^2}{\mathbb{Z}_2}\times \mathbb{C}$ and the field theories has less flavour global symmetries.

\section{Brief discussion on the holographic dual}\label{cf}

In Section \ref{3} we have shown that, assuming the orientifold is able to change the geometry by transforming the CY$_{\mathrm{A}}$ cone into the CY$_{\mathrm{B}}$ cone we are forced to consider relevant operators which induce an RG flow that we know must end in a new superconformal fixed point. In holography we are able to build up gravity solutions that are dual to RG flows \cite{Ammon:2015wua}.  

\subsection{The toy model for holographic duals of RG flows and beyond}\label{4}
Let us consider a set of scalar fields $\phi_a$ in a AdS$_5$ background with coordinates $(x^0,x^1,x^2,x^3,z)$\footnote{We are going to use Greek indices for $4$-dimensional quantities and capital Latin indices for $5$-dimensional ones} and action given by
\begin{equation}
    S=\int d^4xdz \sqrt{-g}\bigg[\frac{\mathcal{R}}{4}-\frac{1}{2}G_{ab}\partial_A\phi_a\partial^A\phi_b-V(\phi)\bigg].
    \label{eqmotionrg}
\end{equation}
We assume a $4D$ Poincaré invariant metric of the form $ds=dz^2+\mathrm{e}^{2Y(z)}dx_{\mu}dx^{\mu}$, known as the domain wall ansatz, that we recognize as Poincaré patch AdS with radius $R$ if the warp factor is a linear function $Y(z)=\frac{z}{R}$ and using the redefinition $\mathrm{e}^{\frac{z}{R}}=\frac{R}{z}$. Fields equations read
\begin{equation}
\begin{aligned}
&3(\partial_zY)^2-\frac{1}{2}G_{ab}\partial_z\phi_a\partial_z\phi_b+V=0;\\
&G_{ab}\partial_z^2\phi_b+4\partial_zYG_{ab}\partial_z\phi_b=\frac{\partial V}{\partial \phi_a};
\end{aligned}    
\end{equation}
an obvious solution is given by the negative critical points of the potential $V_{crit}$ and a set of constant scalars,
\begin{equation}
    \frac{\partial V}{\partial \phi_a}=0, \ \ \ \partial_z\phi_a=0;
\end{equation}
then the second equation is trivially satisfied while the first gives us $(\partial_zY)^2=-\frac{V_{crit}}{3}$. Up to coordinates redefinition we can then write 
\begin{equation}
    Y(z)=\frac{z}{R} \ \ \ with \ \  \frac{1}{R^2}=-\frac{V_{crit}}{3}
\end{equation}
this is nothing but AdS with its radius controlled by the critical point of the potential. Now we can expand the action around this solution and we can infer the conformal dimension of a CFT operator $O_a$ from the masses of the quadratic fluctuations $m_a$,
\begin{equation}
    R^2m_a^2=\Delta_a(\Delta_a-4).
\end{equation}

At this point we are looking for more generic solutions that are asymptotically AdS; we will consider solutions which have linear warp factor $Y(z)=\frac{z}{R}$ and constant $\phi_a$ near the boundary at $z \rightarrow \infty$ and in the deep interior for $z \rightarrow -\infty$. This is conjectured to be dual to an RG flow from a UV fixed point to an IR fixed point. It is natural
to identify the radial coordinate $z$ with the field theory RG scale via $\mu=\mu_0\mathrm{e}^{\frac{z}{R}}$, this choice guarantees that in the UV, at the AdS boundary, we have $\mu \rightarrow \infty$ for $z \rightarrow \infty$, while in the deep interior we have $\mu \rightarrow 0$ for $z \rightarrow -\infty$. As we know the exact identification of the RG scale is scheme dependent and a particular choice of coordinates on the supergravity side corresponds to a particular choice of renormalisation scheme on the field theory side. Said in other way, our goal is to find an interpolating flow solution of (\ref{eqmotionrg}) which interpolates between two stationary points. In AdS/CFT language, this means that we are looking for a domain wall solution interpolating between an AdS space of radius $R_{UV}$ for $z \rightarrow \infty$ and another AdS space of radius $R_{IR}$ for $z \rightarrow -\infty$. At the same time, the scalars $\phi_a$ are expected to flow from a constant $\phi_{a_{UV}}$ in the UV to a constant $\phi_{a_{IR}}$ in the IR. A domain wall solution of this type is expected to be dual to a field theory RG flow between two conformal theories. \\
Since for $z \rightarrow \pm\infty$ we have an AdS space-time we can use the holographic dictionary: on the boundary and in the deep interior region the scalars behave as 
\begin{equation}
    \phi_a(z)\simeq A_az^{4-\Delta_a}+B_az^{\Delta_a}=A_ae^{(\Delta_a-4)z}+B_ae^{-\Delta_a z}
\end{equation}
in which $A_a$ is the source of $O_a$ and $B_a$ is its VEV. If $A_a \neq 0$ the solution describes a deformation of the CFT by operator $O_a$ while if $A_a=0$ and $B_a \neq 0$ it describes a different vacuum of the CFT where the operator $O_a$ develops a non vanishing VEV. In both cases conformal invariance is broken and a RG flow is triggered; the gravity solution is dual to this RG flow. \\
Particularly interesting is the case in which the $O_a$ operator is relevant and the RG flow leads to a new fixed point and this is possible if and only if the potential $V(\phi)$ has more than one critical point (in such a way that another AdS solution with different radius is possible); the gravity solution is a kink interpolating between the two critical point AdS solutions. In order to have an RG flow that starts from a CFT$_\mathrm{UV}$ we need a relevant operator (very similar to what happens in IR duality linked to the algebro-geometrical orientifold) and to hit a CFT$_\mathrm{IR}$ we need the operator to become irrelevant there. Looking at the mass-dimension formula this means that in the relevant operator case the squared mass of the scalar fluctuation must be negative (a maximum of the potential $V(\phi)$) while in the irrelevant operator case the squared mass must be positive (a minimum of the potential $V(\phi)$).

The construction of holographic dual of RG flows could be applied to explain 
the IR duality arising from algebro-geometrical orientifold. We have seen how the duality requires the presence of relevant operators that deform the initial CFT associated to $\mathrm{CY_A}$ cone into the CFT associated to $\mathrm{CY_B}$ cone. In the spirit of the toy model presented above, this entire IR duality mechanism could be interpreted as the holographic dual of a suitable effective supergravity solution that has to interpolate between the supergravity solution of oriented model $\mathrm{A}$ and the supergravity solution of unoriented model $\mathrm{B}$. Let us label the oriented models as $o\mathrm{CFT_A}$ and $o\mathrm{CFT_B}$ while the unoriented models as $u\mathrm{CFT_A}$ and $u\mathrm{CFT_B}$. If we are able to perform the Kaluza-Klein (KK) reduction on the SE base manifolds $X^5_{\mathrm{A}}$ and $X^5_{\mathrm{B}}$ we get two gauged supergravity descriptions with two different potentials $V_{X^5_{\mathrm{A}}}$ and $V_{X^5_{\mathrm{B}}}$.
We expect these two gauged supergavity theories are consistent truncations of type $\mathrm{IIB}$ superstring theory, respectively, on $AdS_5 \times X^5_{\mathrm{A}}$ with gauge group $G_{X^5_{\mathrm{A}}}$ and on $AdS_5 \times X^5_{\mathrm{B}}$ gauge group $G_{X^5_{\mathrm{B}}}$. 
This is in analogy with the FGPW flow \cite{Freedman:1999gp}, which is the holographic description of the Leigh-Strassler RG flow \cite{Leigh:1995ep} of $\mathcal{N}=4$ SYM theory.
Moreover, as in the FGPW flow and according to AdS/CFT philosophy, we expect the gauge groups are given by the isometry groups of the SE bases. \\
The flow would be holographically described if we are able to find a suitable supergravity description with a potential which has several critical points. The maximum of the potential has to preserve the original $G_{X^5_{\mathrm{A}}}$ symmetry while another critical point has to preserve the $G_{X^5_{\mathrm{B}}}$ symmetry. However, this kind of approach poses technical and conceptual challenges. First of all the KK reduction on the compact Sasaki-Einstein manifold $X^5$ is far more difficult than for $S^5$. Therefore, it is not easy at all to build up explicit potentials for these supergravity descriptions. Moreover, from the conceptual point of view, the supergravity description cannot be the full story since the orientifold is non-perturbative. The supergravity description has to be uplifted in a full string theory description which must match with the string theories associated with the two CFTs $\mathrm{A}$ and $\mathrm{B}$. Anyway, it is not clear if such suitable supergravity model exists but if so, we will have a holographic description of the flow induced by the algebro-geometrical orientifold itself and further investigation on this point will be part of future research works.

\section{Conclusion}
Orientifold projections give the opportunity of new possibilities to the creation of field theory models from string background called unoriented models \cite{Dudas:2006bj}. The appellation "unoriented" is mainly due to the action of the orientifold map on the string oscillators: the involution makes oriented string unoriented one and a Chan-Paton like analysis suggests the possibility of new gauge group like $SO(N)$ and $Sp(2N)$. However, orientifold behaviour is inferred by comparison with the flat-space behavior but, probably, orientifold have an intrinsic non-perturbative nature \cite{Cordova:2019cvf}. This is also suggested by a recent interesting new feature proposed in \cite{2020in} and expanded in \cite{2022su}: the so-called scenario III. What happens is that a theory after a scenario III orientifold shares the same $R$-charges, central charge and superconformal index of another orientifolded theory coming from a scenario I orientifold. The crucial point is that these physical quantities are the same not only in the large $N$ limit but for every $N$ \cite{2020in,2022su}: this is a non-perturbative effect. Since the physical quantities are the same in the IR we refer to these theories as IR duals and, since in the geometrical engineering the field theory depends on the geometry of the CY cone, these two orientifolded theories can be considered to have the same CY cone geometry. This means that a changing in the geometry of the cone has to happen due to the orientifold action. The change, due to the orientifold, can be traced back to the equations that describe the algebraic CY varieties and we require the matching between these sets of equations. In this sense the orientifold has an algebro-geometrical interpretation as a morphism between algebraic varieties given locally by polynomials. This interpretation has as its guiding idea the flop transition and the other geometrical changing in string theory compactifications \cite{Greene:1996cy,Aspinwall:1993nu,Greene:1995hu,Witten:1993yc}. The properties and a more deep understanding of this interpretation, such as the answers to the questions: "Are orientifold regular maps or rational maps? Are they biregular or birational?" could be fertile ground for future works. We stress that if orientifold are such kind of maps for the CY cones they open the door to the use of the theory of category with orientifold as morphisms and field theory as objects.
Therefore, orientifold acts both on string states and on the geometry of the space-time. 
\begin{figure}[h]
    \centering
    {\includegraphics[width=.80\textwidth]{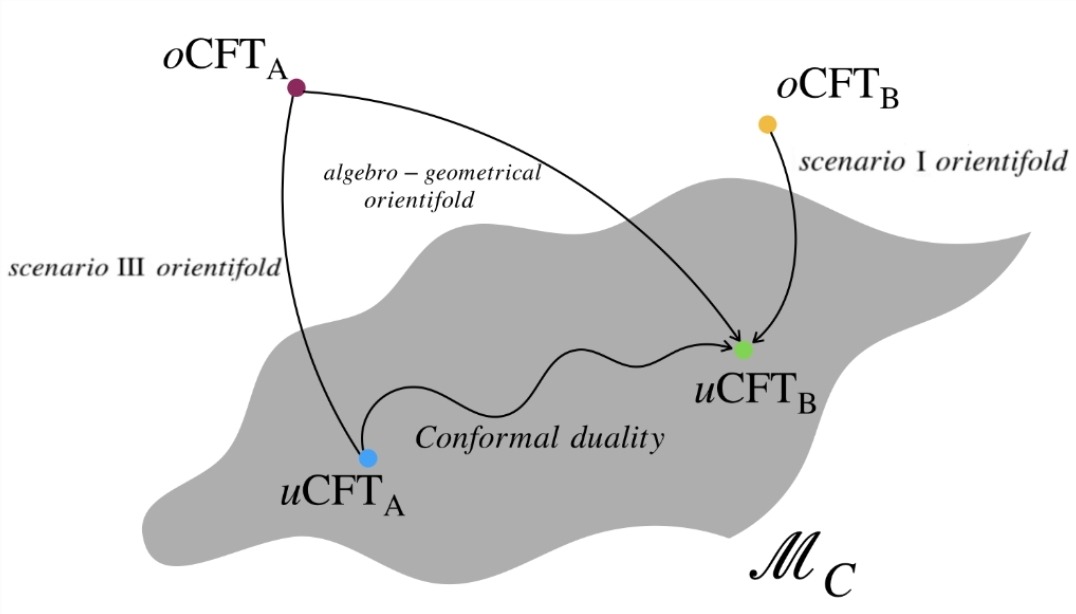}}
\caption{\textit{Schematic picture of the RG flow triggered by the orientifold action. The little "o" stays for "oriented" while the little "u" stays for "unoriented". The two unoriented models lie on the same conformal manifold while the two oriented models are not related in any way. The flow triggered by the algebro-geometrical orientifold action links the oriented model $\mathrm{A}$ to the unoriented model $\mathrm{B}$.}}
\label{fig7}
\end{figure}

From the requirement of the matching between these sets of equations describing the CY cone of the two theories and using some tools of algebraic geometry and toric geometry, reported in \ref{2}, we infer the existence of relevant operators that deform the initial CFT triggering an RG flow towards the IR regime. A schematic picture of the flow is given in Figure \ref{fig7}. 
The flow could be described using a holographic dual, whose toy model is presented in \ref{4}, but the prohibitive Kaluza-Klein reduction on the base of a generic CY cone makes difficult the explicit construction. Moreover, due to the intrinsic non-perturbative nature of orientifold, this supergravity dual is expected to be only an effective description that must necessarily be uplifted in full string theory.

\newpage

\subsubsection*{Acknowledgment}
We thank first of all Prof. Fabio Riccioni to have inspired the present work and Prof. Massimo Bianchi and Salvo Mancani for interesting discussions. We also thank Davide Morgante for discussions and comments and Matteo Romoli for its reading of the manuscript and comments. Special thanks to Maria Luisa Limongi who supported the work.

\bibliographystyle{JHEP} 
\bibliography{biblio}

\end{document}